\documentclass[a4paper]{jpconf}
\usepackage{graphicx}
\usepackage{xspace}

\newcommand{\UNIT}[1]{\ensuremath{\,{\rm #1}}\xspace}
\newcommand{\GeV}{\UNIT{GeV}}

\begin{document}
\title{Dilepton Production in Transport-based Approaches}

\author{Janus Weil$^1$, Stephan Endres$^1$, Hendrik van Hees$^1$, 
Marcus Bleicher$^1$, Ulrich Mosel$^2$}

\address{$^1$Frankfurt Institute for Advanced Studies , Ruth-Moufang-Str.~1, 60438 
Frankfurt, Germany\\
$^2$Institut f\"ur Theoretische Physik, JLU Giessen, Heinrich-Buff-Ring 16, 
35392 Giessen, Germany}

\ead{weil@fias.uni-frankfurt.de}

\begin{abstract}
We investigate dilepton production in transport-based approaches and show that 
the baryon couplings of the $\rho$ meson represent the most important 
ingredient for understanding the measured dilepton spectra. At SIS energies,
the baryon resonances naturally play a major role and affect 
already the vacuum spectra via Dalitz-like contributions, which can be captured 
well in transport simulations. Recent pion-beam measurements at GSI will
help to constrain the properties of the involved resonances further.
\end{abstract}

\section{Introduction}

Lepton pairs are an ideal probe for studying phenomena at high 
densities and temperatures. They are created at all stages of a heavy-ion 
collision, but unlike hadrons they can escape the hot and dense zone almost 
undisturbed (since they only interact electromagnetically) and thus can carry 
genuine in-medium information out to the detector. Dileptons are particularly 
well-suited to study the in-medium properties of vector mesons, since the 
latter can directly convert into a virtual photon, and thus a lepton pair 
\cite{Leupold:2009kz,Rapp:2009yu}.
One of the groundbreaking experiments in this field was NA60 at the CERN SPS, 
which revealed that the $\rho$ spectral function is strongly broadened in the 
medium. Calculations by Rapp et al.~have shown that this collisional broadening 
is mostly driven by baryonic effects, i.e., the coupling of the $\rho$ meson to 
baryon resonances ($N^*$, $\Delta^*$) \cite{vanHees:2006ng}.
In the low-energy regime, the data taken by the DLS detector have puzzled 
theorists for years and have recently been confirmed and extended by new 
measurements by the HADES collaboration 
\cite{HADES:2011ab,Agakishiev:2011vf}. At such low energies, it is expected that 
not only the in-medium properties are determined by baryonic effects, but that 
already the production mechanism of vector mesons is dominated by the coupling 
to baryons (even in the vacuum).


\section{The model: hadronic transport + VMD}

\label{sec:VMD}

Already our previous investigations \cite{Weil:2012ji} based on the GiBUU 
transport model \cite{Buss:2011mx} have shown that the baryonic $N^*$ and 
$\Delta^*$ resonances give important Dalitz-like contributions to dilepton spectra at 
SIS energies for pp as well as AA collisions.
This finding is based on the assumption that these resonances decay into a 
lepton pair exclusively via an intermediate $\rho$ meson (i.e., strict 
vector-meson dominance).
In the transport simulation, the Dalitz decays $R\rightarrow e^+e^-N$ are 
treated as a two-step process, where the first part is an $R\rightarrow\rho N$ 
decay, followed by a subsequent conversion of the $\rho$ into a lepton pair 
($\rho\rightarrow e^+e^-$).
The branching ratios for the $R\rightarrow\rho N$ decays are taken from the 
partial-wave analysis by Manley et al.~\cite{Manley:1992yb},
while the decay width for the second part is calculated under the strict-VMD 
assumption. Recently we have extended the VMD assumption also to the $\Delta(1232)$ 
state \cite{Weil:2014lma}.

In principle it is clear that all the dilepton Dalitz decays of the baryonic
resonances ($R\rightarrow Ne^+e^-$) must involve an electromagnetic transition
form factor. From an experimental point of view this form factor is essentially unknown in the time-like region.
In consequence it is frequently neglected in transport
simulations. Our approach is to use strict VMD as a sensible ansatz for this form factor,
which is easy to implement in a transport scheme.


\section{Dilepton spectra from p+p and A+A collisions}

\begin{figure}[ht]
 \begin{center}
  \includegraphics[width=0.45\textwidth]{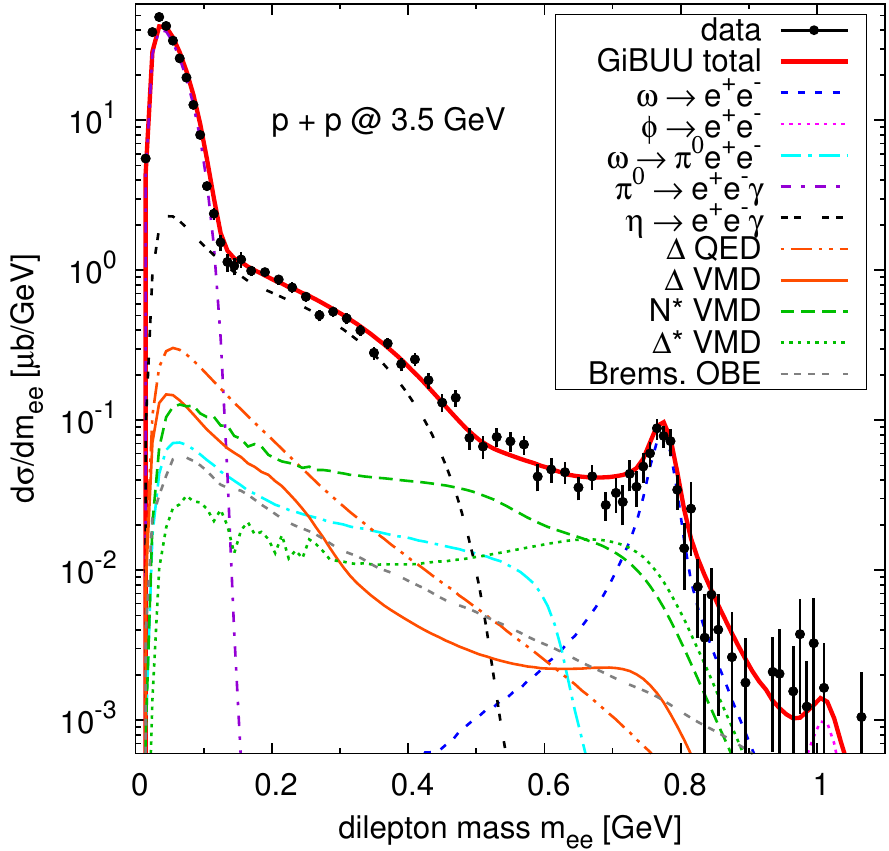}
  \includegraphics[width=0.45\textwidth]{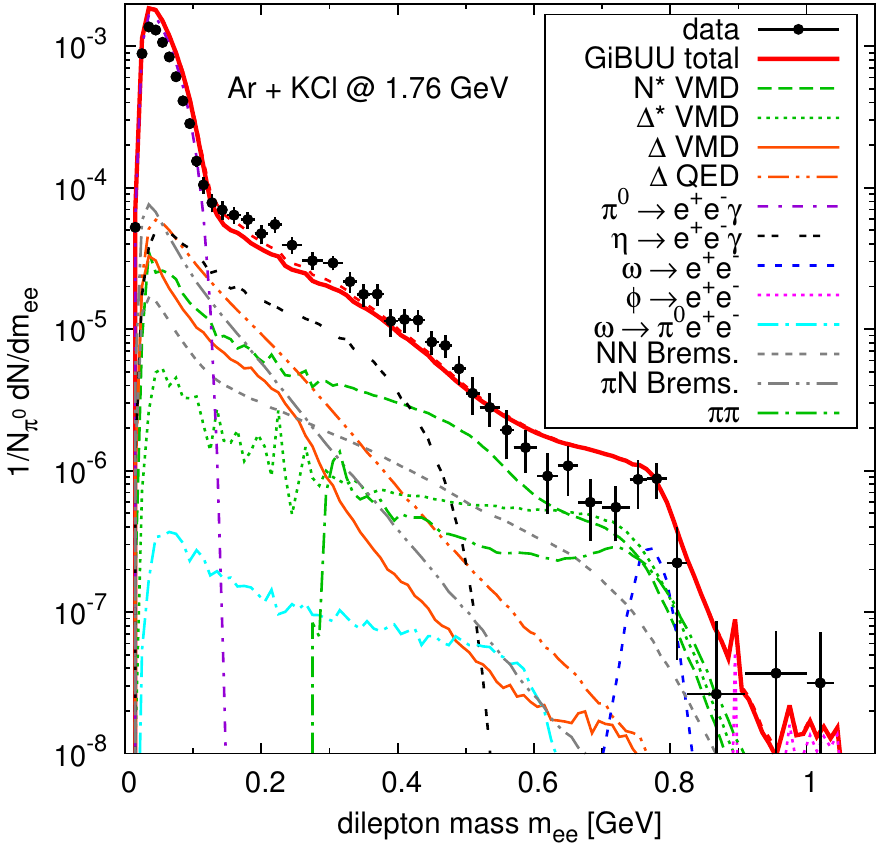}
 \end{center}
  \caption{Dilepton mass spectra for p+p and Ar+KCl collisions, in comparison to the data 
from \cite{HADES:2011ab,Agakishiev:2011vf}.}
  \label{fig:pp}
\end{figure}

In Fig.~1 we present two examples of results obtained with our model \cite{Weil:2014lma}.
The left-hand side shows the simulated dilepton mass spectrum from elementary p+p
collisions at $E_{\rm kin}=3.5\GeV$ (inside the detector acceptance) compared to the
data measured by the HADES collaboration.

While the low-mass regime is dominated by the Dalitz decays of the $\pi^0$ and
$\eta$ mesons, the region of higher masses shows a clear $\omega$ peak (whose
width is dictated by detector resolution) and a hint of a $\phi$ peak at the very end
of the spectrum. The most interesting region, however, is the intermediate-mass
regime below the $\omega$ peak, which according to our model is dominated by
baryonic contributions from $N^*$ (and $\Delta^*$) resonances. There is
a whole cocktail of such resonances which are known to couple to the $\rho$ meson,
but in particular the light ones, such as the $D_{13}(1520)$, give important contributions here,
without which the data in the intermediate-mass regime could not be described.
The $\Delta(1232)$ only gives subdominant contributions in our model (both in the QED and VMD approaches),
which is in conflict with other results \cite{Bratkovskaya:2013vx}, which claim a much more dominant role of the $\Delta$.

The right panel of Fig.~1 shows the dilepton spectrum from Ar+KCl collisions at an
energy of 1.76 AGeV compared to the HADES data.
There are additional channels in this medium-sized nucleus-nucleus system, which do not occur in pure pp collisions, such
as $\pi N$ bremsstrahlung or the contribution from $\pi\pi\rightarrow\rho\rightarrow e^+e^-$, but they
are not dominant. As in the elementary case, the contributions from the baryonic
resonances are significant.
Overall we observe an underestimation at intermediate masses and a slight excess in the vector-meson pole region.
Reasons for these deviations could be uncertainties in the used resonance parameters
or a missing dynamical treatment of density-dependent spectral functions.
Such a treatment may be provided by the so-called ``coarse-graining`` 
approach, which is subject of ongoing investigations \cite{Endres:2013cza}.
It could help to improve the description of heavy systems
and to make the connection to dilepton measurements at higher energies.


\section{Dilepton spectra from pion-induced collisions}

\begin{figure}[ht]
 \centering
  \includegraphics[width=0.45\textwidth]{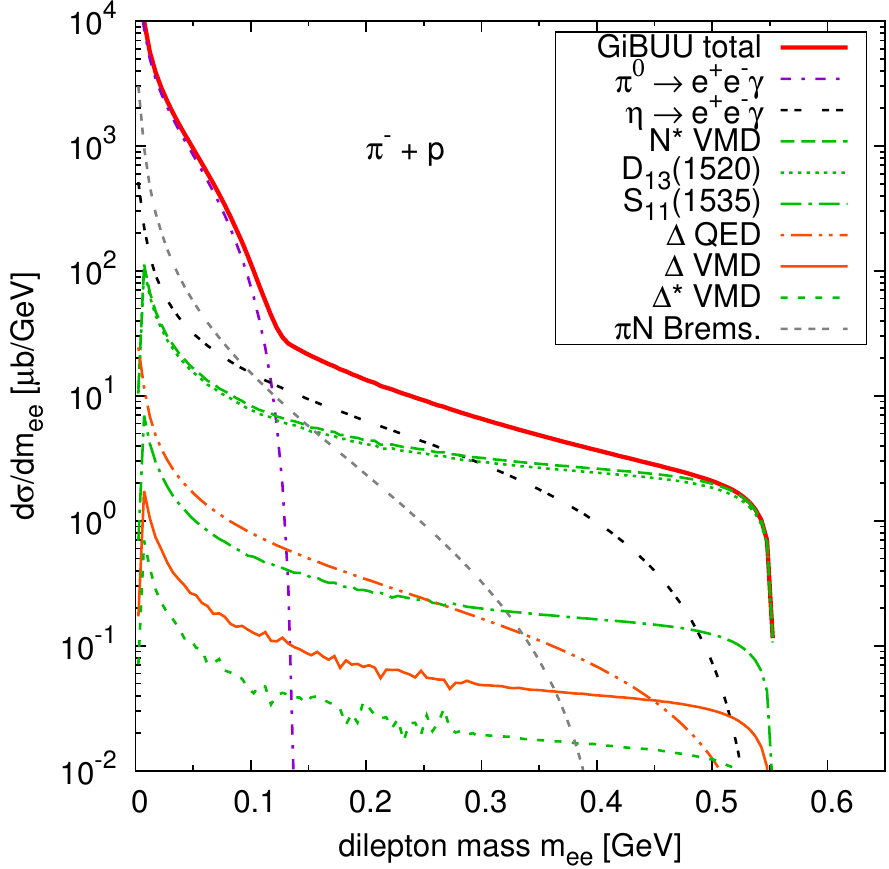}
  \includegraphics[width=0.45\textwidth]{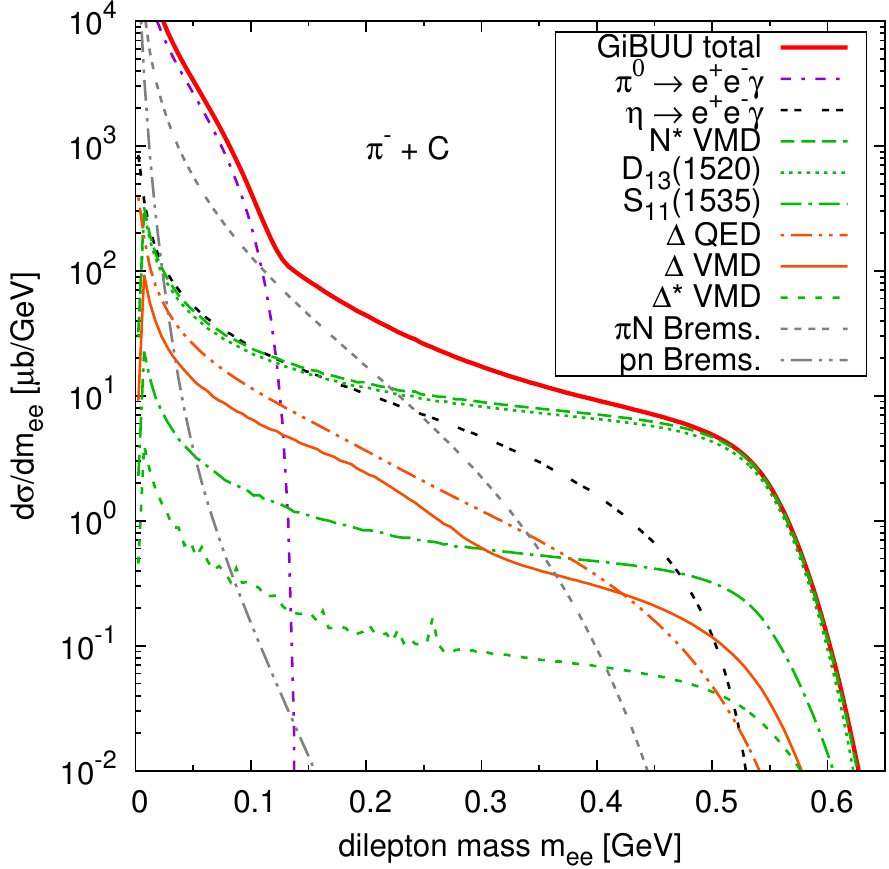}
  \caption{Dilepton mass spectra for $\pi^-p$ and $\pi^-C$ collisions at $\sqrt{s_{\pi N}}=1.49\GeV$.}
  \label{fig:AA}
\end{figure}

Pion-induced reactions offer a much cleaner testing ground for dilepton production
than AA or pp reactions, since the latter always involve excitation of baryonic
resonances at various energies, while $\pi N$ collisions operate
at a fixed $\sqrt{s_{\pi N}}$, corresponding to a fixed resonance mass. This makes
it much easier to disentangle contributions from different resonances, as pointed out
e.g. in \cite{Diss}.

Fig.~2 shows the prediction of our model for dilepton mass spectra from pion-induced
reactions on proton and carbon targets at a beam kinetic energy of $E_{\rm kin}=0.566\GeV$ (corresponding to
$p_{\rm lab}=0.690\GeV$ or $\sqrt{s}=1.49\GeV$) which have recently been measured at GSI.
In contrast to Fig.~1, no detector acceptance and resolution effects are included here.

The considered energy is just above the $\eta$ threshold and
close to the pole masses of resonance states like $D_{13}(1520)$ and $S_{11}(1535)$.
In particular the former is believed to have a strong coupling to the $\rho$ meson
and thus should have a large influence on dilepton spectra. In our model, the $D_{13}(1520)$
plays a very dominant role at this energy. When disregarding the $\eta$ Dalitz contribution,
which can be subtracted in the experimental spectra, essentially all of the
high-mass yield is due to the $D_{13}(1520)$, while contributions from other
resonances like the $S_{11}(1535)$ or the $\Delta(1232)$ are suppressed by roughly
an order of magnitude in the elementary $\pi^-p$ reaction. The $\Delta$ contribution
is again shown in two different approaches, and there also is a $\pi N$
bremsstrahlung contribution. However, the latter should be taken with a grain of salt,
since it is only calculated in soft-photon approximation, which may not be very
reliable in the $\pi N$ case, in particular for large dilepton masses.

When moving to the carbon target, as shown on the right-hand side of Fig.~2, one can
see that the $\Delta$ and bremsstrahlung contributions gain additional strength from
secondary collisions. Yet, the relative strength of the $\eta$ contribution drops a bit,
since it is produced only in the primary collisions, but can be absorbed afterwards.
The dominance of the $D_{13}(1520)$ in the high-mass regions persists.

We stress that our model currently relies on the assumption of a strict-VMD
form factor for the $D_{13}(1520)$ (as for all other $N^*$ resonances). However,
the strong dominance of this resonance in the reaction considered here might
make it possible to obtain a rather direct measurement of its electromagnetic
transition form factor. It will be very interesting to see whether the experimental
data will reveal any deviation from the VMD assumption.

In any case, it should also be mentioned that some models claim rather strong
destructive interference effects at this energy \cite{Lutz:2002nx} (which are neglected in our transport
simulation, since they are very hard to handle). However, those effects are subject
to controversy and do not show up so strongly at this energy in other models \cite{Zetenyi:2012hg}.


\section{Conclusions}

We have shown that the HADES dilepton data from pp and AA collisions can be 
described rather well with a combination of a resonance-model-based transport 
approach with a strict-VMD coupling of the baryons to the electromagnetic sector, where a 
mix of different baryonic resonances contributes to the total dilepton yield.
Unfortunately not all parameters of these resonances are known sufficiently well.
However, recent pion-beam measurements at GSI could help to shed new light on the
baryonic resonances and possibly even fix the transition form factor of their
electromagnetic Dalitz decays.

\section*{Acknowledgments}

This work was supported by the Hessian Initiative for Excellence (LOEWE) 
through the Helmholtz International Center for FAIR and by the Federal Ministry 
of Education and Research (BMBF). J.W.~acknowledges funding of a Helmholtz 
Young Investigator Group VH-NG-822 from the Helmholtz Association and GSI.

\end{document}